# Dynamic emergence of fragmenting, exploding cages of irradiated single-walled carbon nanotubes and $C_{60}$

Shoaib Ahmad
National Center for Physics, QAU Campus, Shahdara Valley, Islamabad 44000, Pakistan

sahmad.ncp@gmail.com

## Abstract

Experimental results from irradiated-carbon nanostructures can be partially explained by selective applications of classical damage theories and ad-hoc thermal models by treating the binary atomic collision cascades and multi-atomic thermal spikes as energy-dissipating mechanisms. An information-theoretic model is developed by treating the irradiated single-walled carbon nanotubes and $C_{60}$ as entropy-generating dissipative structures. The model is based on evaluating the experimental probability distribution functions of the sputtered constituents and components from the irradiated carbon nanostructures that yield their Shannon entropy or information. Three information-based functions of fractal dimension, relative entropy and dynamic emergence are defined and employed to characterize the profiles of self-organizing, information-generating dissipative structures. Fractal dimension determines the space-filling character, relative entropy establishes the distance between any two of the dissipative structures. Dynamic emergence function provides the profile of the emerging sequences. Results derived from the existing theories and thermal models are compared with the information-theoretic dynamic emergence model to comprehensively describe the nature of the dissipative structures in irradiated carbon nanostructures without making *apriori* assumptions about the energy-dissipating mechanisms.

# Introduction

Solids irradiated by external energetic ions can be treated as dynamical systems[1-3] where bifurcating recoil atomic trajectories may lead to collision cascades (CC) or localized thermal spikes (LTS). The ensuing disorder in the form of point, line and volume defects produces vacancies, sputtered atoms, dislocations and clusters of defects. Dynamics of the emergence of these defects with specific temporal and spatial profiles characterize these as entropy-generating dissipative structures (DS).[4-7] Non-equilibrium binary scattering mechanisms for CC are described by sputtering theory and Monte Carlo simulations in 3D bulk solids.[8,9]. Results from irradiated carbon nanostructures[10-13] have demonstrated the need for revisiting the theoretical framework to explain the irradiation-induced damage on 2D surfaces of single-walled carbon nanotubes (SWCNTs) and $C_{60}$ cages. Investigations of the simultaneous emergence of CC and LTS in irradiated SWCNTs have been treated as transition of non-equilibrium cascades to localized hotspots-LTS.[7] Thermal models and simulations[14-17] yield atoms and cluster dissociation temperatures in SWCNTs, but do not present the dynamic profile of the cage-to-constituents explosive transition. An information-theoretic model is presented here for the dynamic emergence of CC, LTS, fragmenting and exploding cages as self-organizing dissipative structures.[18-20] This model is firmly rooted in Shannon's information-theoretic paradigm[21-27] where the channel is the reservoir of emerging, information-generating DS.[28,29] Information is derived from the probability distribution functions of the sputtered atoms and clusters emitted from the irradiated SWCNTs and $C_{60}$s as a function of irradiating ion energy. Information and Kullback-Leibler distance[26] or relative entropy between two probability distributions, fractal dimension[30-36] and a new dynamic emergence function is introduced that yields the emerging trend of interacting, competing, evolving and disintegrating DS of irradiated SWCNTs and $C_{60}$s.

In this communication we use CC≡collision cascades, LTS≡localized thermal spikes and DS≡dissipative structures.

## Methods

Probability distribution functions $p(C_x) \equiv p_x(\zeta)$ for the sputtered anions $C_x$ are evaluated from mass spectra as a function of cesium energy $E(Cs^+)$. SWCNTs and $C_{60}$s are irradiated in Source of Negative Ions with Cesium Sputtering-SNICS.[37] SWCNTs of nominal diameter $\sim 3$ nm, average length $\sim 8-10$ microns were compressed in Cu bullet targets for SNICS installed on 2 MV Pelletron at GCU, Lahore. The. $Cs^+$-sputtered atoms and clusters are extracted as anions. A 30º momentum analyzer delivered mass spectra of the sputtered anions as a function of cesium energy $E(Cs^+)$. The experiments were conducted by defining the $Cs^+$ energy range $E(Cs^+)$ and choosing the appropriate scale of the incremental energy steps $\delta E(Cs^+)$. Scale variation of $\delta E(Cs^+)$ provides the experimental measure $\zeta = \zeta(E(Cs^+), \delta E(Cs^+))$. Normalized yields $N_{C_x}$ for $C_x$ with $x$-atoms obtained from each mass spectrum provide probabilities of emission $p(C_x) \equiv N_{C_x}/\Sigma N_{C_x}$. The probability distribution function $p(C_x) = p_x(E(Cs^+), \delta E(Cs^+)) \equiv p_x(\zeta)$ represents the combined effects of $E(Cs^+)$ and $\delta E(Cs^+)$. In the experiments reported here and the analysis thereafter, $\zeta$ is the basic measure in our experimental data with $\delta E(Cs^+)$ set at 0.1 and 0.2 keV.

## Fragmenting SWCNTs and $C_{60}$ cages

$C_2$ emerges as the main sputtered species rather than $C_1$, while CC-based sputtering theories predict only the monatomic $C_1$ emission.[8,9] The dominance of multi-atomic carbon clusters over $C_1$ from $Cs^+$-irradiated SWCNTs and $C_{60}$ has been amply documented.[11-13] Linear CC do not explain the sputtering of clusters. Cascades of binary atomic collisions create single vacancies and mobile atomic defects with ion-energy dependent probability $p(C_1) \propto E(Cs^+)$.[8,9]

Our experiments show $E(Cs^+)$-independent cluster $p(C_{x>1})$ emissions. Clusters are neither considered nor expected to be sputtered from cascades as bifurcating atomic collisions do not generate multi-atomic vacancies. Clusters has been shown to be the result of the emerging LTS in SWCNTs with emission probabilities $p(C_x) \approx \{exp(E_{xv}/kT_{Sub}) + 1\}^{-1}$, where $E_{xv}$ is binding energy of an *x*-member vacancy at sublimation temperature $T_{sub}$.[14] Simulations with irradiated bulk solids (Si and Ge) have also indicated the emerging nonlinear energy dissipation mechanisms via LTS in addition to CC.[38-41] From the radiation damage perspective, the ensuing damage has two time scales $\tau_{LTS} \gg \tau_{CC}$, CC occur for fractions for ps and LTS lasts for many ns. Both mechanisms have collisional relationship and share the same spatial regions. For times $>\tau_{LTS}$, the damage annealing sequences operate to minimize the localized sublimation-induced damage.[41] This explains the reason that the net irradiation-induced damage at times $\gg \tau_{LTS}$ cannot provide the evolutionary trails of the emergence of the sequences of damage. From the same experimental results of the dynamic evolution of $p(C_x)$ of sputtered species we develop thermal models and the information-theoretic diagnostic tools.

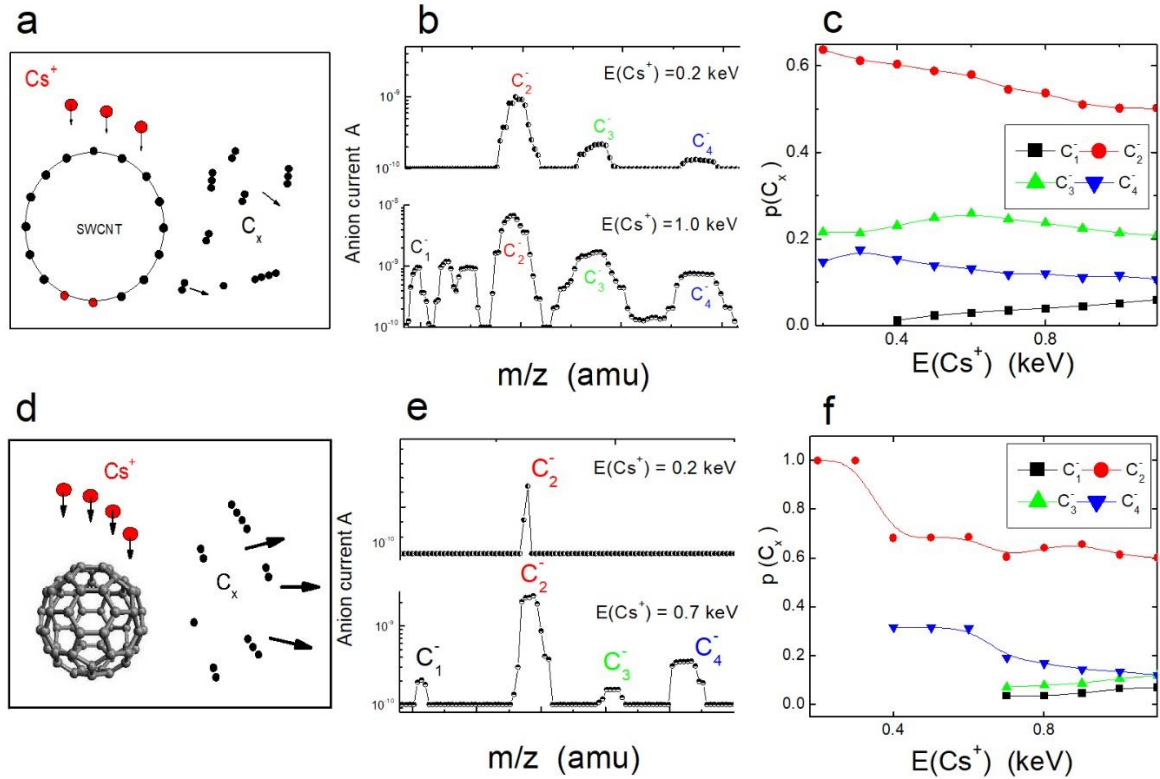

Fig. 1. Irradiated SWCNTs and $C_{60}$-fullerite with $E(Cs^+) = 0.2 - 1.1$ keV, $\delta E(Cs^+) = 0.1\ keV$. (a) Schematics of $Cs^+$-irradiation of SWCNT and sputtering of atoms and clusters. (b) Two mass spectra of sputtered anions at $E(Cs^+)$ = 0.2 and 1.0 keV are shown. Only clusters are emitted at 0.2 keV, while a low intensity $C_1$ anion is noticeable at 1.0 keV. (c) Probability distribution function $p(C_x)$ is evaluated for the four sputtered anions $C_x^-$ from the normalized yields in each mass spectrum as a function of $E(Cs^+)$. (d) Schematic diagram of $Cs^+$-irradiation of $C_{60}$ is shown with emitted atoms and clusters. (e) Mass spectrum at $E(Cs^+)$ = 0.2 keV has only $C_2^-$ is emitted, while all sputtered species $C_1^-$, $C_2^-$, $C_3^-$ and $C_4^-$ are present in the spectrum at $E(Cs^+)$ = 0.7 keV. (f) $p(C_x)$ is constructed for each of the emitted species, from the mass spectra obtained as a function of $E(Cs^+)$.

Figure 1 presents the essential features of experiments results with schematic diagrams in 1(a) and (d) of $Cs^+$-irradiation of a SWCNT and $C_{60}$ with the consequent emissions of atoms ($C_1$) and clusters ($C_2$, $C_3$ and $C_4$). Figure 1(b) has two representative spectra at $E(Cs^+) = 0.2\ keV$ has clusters only, while spectrum at 1.0 keV shows C₁ in addition to clusters. $p(C_x)$ of emitted species are grouped in figure 1(c) for $E(Cs^+) = 0.2 - 1.1 \, keV$. Similar experimental conditions were used for C$_{60}$-fullerite. Two mass spectra in figure 1(e) has C$_2$-spitting C$_{60}$ cages at 0.2 keV, while fragmentation by emitting C$_2$ and C$_4$ with C$_1$ and C$_3$ as minor constituents occurs at $E(Cs^+) = 0.7 \, keV$. In figure 1(f) $p(C_x)$ as a function of $E(Cs^+)$ illustrates the emerging features of the irradiated C$_{60}$ cages' fragmenting sequences, from C$_2$ emissions at $E(Cs^+) = 0.2, 0.3 \, keV$, C$_2$ and C$_4$ emissions at $E(Cs^+) = 0.4 - 0.7 \, keV$ and the C$_1$, C$_2$, C$_3$ and C$_4$ emissions for $E(Cs^+) \geq 0.7 \, keV$. The topological diversity of the static, tubular SWCNTs with a large number of C atoms ($\gg 1$) and the rotating, spherical molecules of C$_{60}$ with fixed number of C atoms are vividly demonstrated in their respective fragmenting patterns shown in figure 1. $C_x$-spitting, irradiated SWCNTs transform into damaged SWCNTs with monatomic and multi-atomic vacancies $\{Cs^+ + (SWCNT)_{pristine}\} \rightarrow \{(SWCNT)_{damaged} + C_{x\geq 1}\}$. Irradiation of C$_{60}$ cages shows sequences of fragmentation leading to explosion of the shrinking cage in the sequences $C_{60} \rightarrow C_{58} + C_2; C_{60} \rightarrow C_{54} + C_2 + C_4 \rightarrow \sum_{x\geq 1} C_x$. The emergence profiles of fragmenting SWCNTs can be described by a thermal model and exploding C$_{60}$ cages will be argued with the help of a kinematical model. Conclusions drawn from the thermal models are compared with a comprehensive information-theoretic SRS model of dissipative structures[28] developed in the next section.

## Information-theoretic diagnostic tools for emerging dissipative structures

Irradiating ion energy $E(Cs^+)$ along with its systematic variations $\delta E(Cs^+)$ are the basic parameters of the emerging dynamical system $\{Cs^+ + (SWCNT, C_{60})\}$ that represents the initial input from the source of $Cs^+$ into the reservoir of sp²-bonded networks of carbon atoms in the form of SWCNT or $C_{60}$. This reservoir is analogous to Shannon's channel[21] with a subtle difference that we treat it as a dynamical system, a repository of material and energy. SRS applied to irradiated SWCNTs and $C_{60}$ in Fig.2. Here information is generated. In terms of Shannon theory, we have a 'noisy-channel'. This 'noisy-channel' is treated as the emerging reservoir (the irradiated-SWCNTs or $C_{60}$) that can generate and manipulate information. Modular description of such an information-theoretic Source-Reservoir-Sink (SRS) model[28] was developed to describe the emergence of $C_{60}$ out of ensembles of self-organizing fullerenes.[29] Here, it is employed, with an additional function defined as dynamic emergence, to investigate the emerging trends of the fragmenting, irradiated SWCNTs and $C_{60}$s. The results of SRS model will be compared with those obtained by evaluating these two irradiated nanostructures by thermal models by utilizing $p(C_x)$ of the emitted carbon atoms and clusters ($C_x; x \geq 1$). SRS model treats such a dynamical system as entropy-generating DS where Shannon entropy or information

$$I_x = \sum p(C_x)\ln(1/p(C_x)) \qquad (1)$$

for each emitted species. The two sp[2]-bonded carbon nanostructures have different topologies; SWCNT is cylindrical while $C_{60}$ is a spherical nano-cage. The other significant difference is the nature of $Cs^+$-target interactions. SWCNT presents a static, hexagon-based nano-structured target where energy dissipation created cascades of binary recoiling atoms. Thermal spikes may originate from the enhanced localized vibrational motion of the atoms of hexagons of SWCNTs. In the case of $C_{60}$, a rotating, vibrating molecule is irradiated. Even in the condensed form of fullerite, where translational motion is restricted, each $C_{60}$ molecule continues to rotate around a randomly oriented axis. We report results from two target configurations, the irradiated $C_{60}$-fullerite and $C_{60}$ molecules suspended in ZnO powder. In the case of irradiated fullerite, the cages fragment sequentially by spitting out $C_2$ and $C_4$ before exploding. Irradiation of $C_{60}$-cages suspended in ZnO powder cage-explosion seems to be the preferred route of ion-energy dissipation and entropy generation.

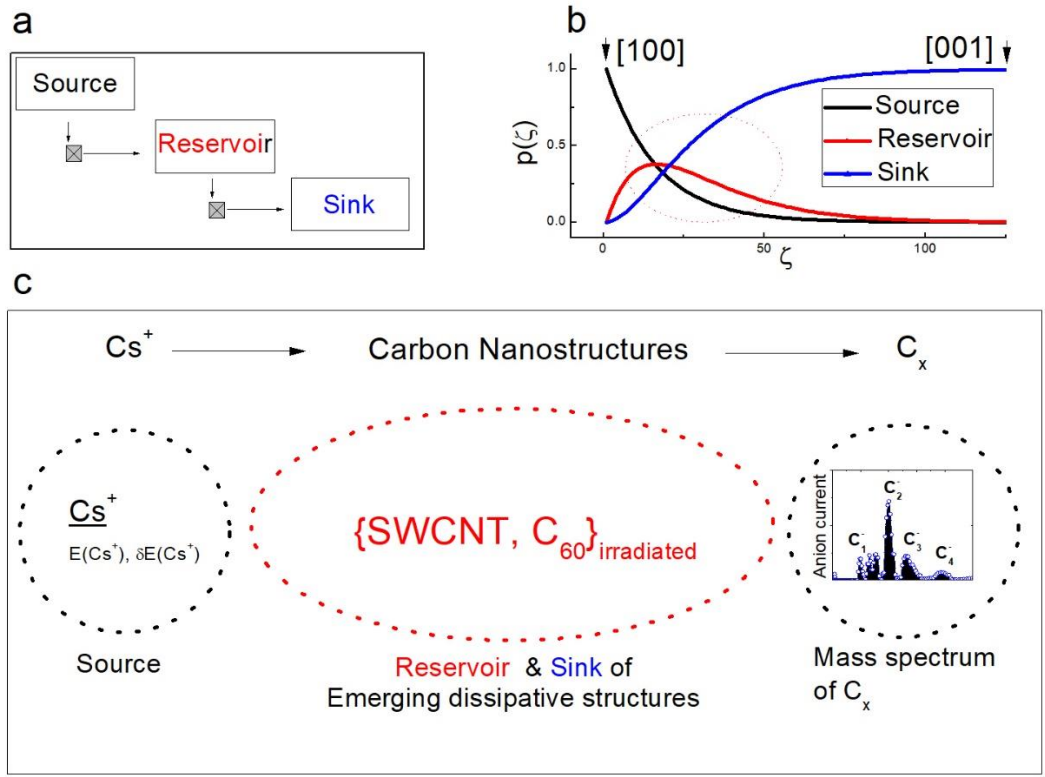

Fig. 2. SRS-model applied to irradiated-SWCNTs and $C_{60}$. (a) Source-Reservoir-Sink (SRS) model is shown as information generating, manipulating and sharing Boxes. (b) Probability distribution function $p(\zeta)$ is constructed for each Box at a given flow-rate of information between the boxes. $p(\zeta)$ is plotted as a function of $\zeta$ which is defined as the number of information-sharing stages, at a given flow-rate. The initial state of SRS is pointed with arrow as [100]. It transforms to [001] for large $\zeta$, i.e., when the information transfer to Sink has completed. The region shown as dotted ellipse is where information-sharing, generating and transformation is happening. (c) The physical representation of SRS is shown where $Cs^+$ (representing Source) irradiates carbon nanostructures with the consequent emission of $C_x^-$. Reservoir to Sink transformation is represented as $\{Cs^+ + (SWCNT, C_{60})_{pristine}\} \rightarrow \{(SWCNT, C_{60})_{damaged} + C_{x \geq 1}\}$. Mass spectrum of the emitted $C_x^-$ carries signatures of the emerging dissipative structures that yield experimental $p(C_x)$.

The emerging DS are diagnosed with Renyi's definition of fractal dimension[25]

$$d_f^x = I_x/\ln(1/\zeta) \qquad (2)$$

here $\zeta$ is the measure of scale. $d_f^x$ is calculated for every $C_x$ from its information $I_x$. It will be shown that $d_f^x$ emerges as a DS-defining function. It helps to distinguish linear versus nonlinear physical mechanisms. Kullback-Leibler distance[26] or the relative entropy[27] for any two probability distributions $p_x(\zeta)$ and $p_y(\zeta)$ of the sputtered species as

$$D(p_x \parallel p_y) = \sum p_x(\zeta) ln(p_x/p_y) \qquad (3).$$

It is a measure that will be used to confirm the nature of the irradiation-induced, diverse physical processes identified by their respective fractal dimensions. For example, if monatomic emissions signal a linear CC and $C_2$ and $C_3$ are considered the flag-bearers of nonlinear LTS then $D(C_2 \parallel C_1) > D(C_2 \parallel C_3)$. This would imply that Kullback-Leibler distance between the probability distributions of $C_2$ and $C_1$ is larger than that between $C_2$ and $C_3$ implying two distinctly different physical mechanisms for the emissions of $C_2$ and $C_1$. Dynamic emergence $\varepsilon_x$ evaluated for all of the sputtered species $C_x$ as a function of $\zeta$, provides comprehensive profiles of the emerging DS. Dynamic emergence function

$$\varepsilon_x = p_x \ln(p_x)/\sum_x p_x \ln(p_x) \qquad (4)$$

$\varepsilon_x$ and its derivative $d\varepsilon_x/d\zeta$ are used to map the evolution of diverse ion-induced structural transformations as a function of $\zeta$. Information-theoretic diagnostic tools of $I_x, d_f^x, D(p_x \parallel p_y)$ and $\varepsilon_x$ provide coherent trends of the emerging and evolving DS without *apriori* assumptions about the physical mechanisms.

## Dynamic emergence and the dissipative structural transformations

Figures 3(a), (c) and (e) are three sets of experimental data of for $p_x(\zeta)$ as a function of $\zeta$. SWCNTs and $C_{60}$. All data points are plotted as a function of $\zeta$ for each value of $E(Cs^+)$ at the chosen scale $\delta E(Cs^+)$. Same energy range $E(Cs^+) = 0.2 - 2.0\ keV$ and $\delta E(Cs^+) = 0.1\ keV$ is used in Fig. 3(a) and (c) where $p_x(\zeta)$ versus $\zeta$ are plotted for SWCNTs and $C_{60}$-fullertie. For $C_{60}$-ZnO powder $E(Cs^+) = 0.8 - 5.0\ keV$, $\delta E(Cs^+) = 0.2\ keV$ in figure 3(e) for $p_x(\zeta)$ of the emitted species. Dynamic emergence $\varepsilon_x$ for $C_1$-$C_4$ derived from data in 3(a), (c) and (e) are plotted in figures 3(b), (d) and (f).

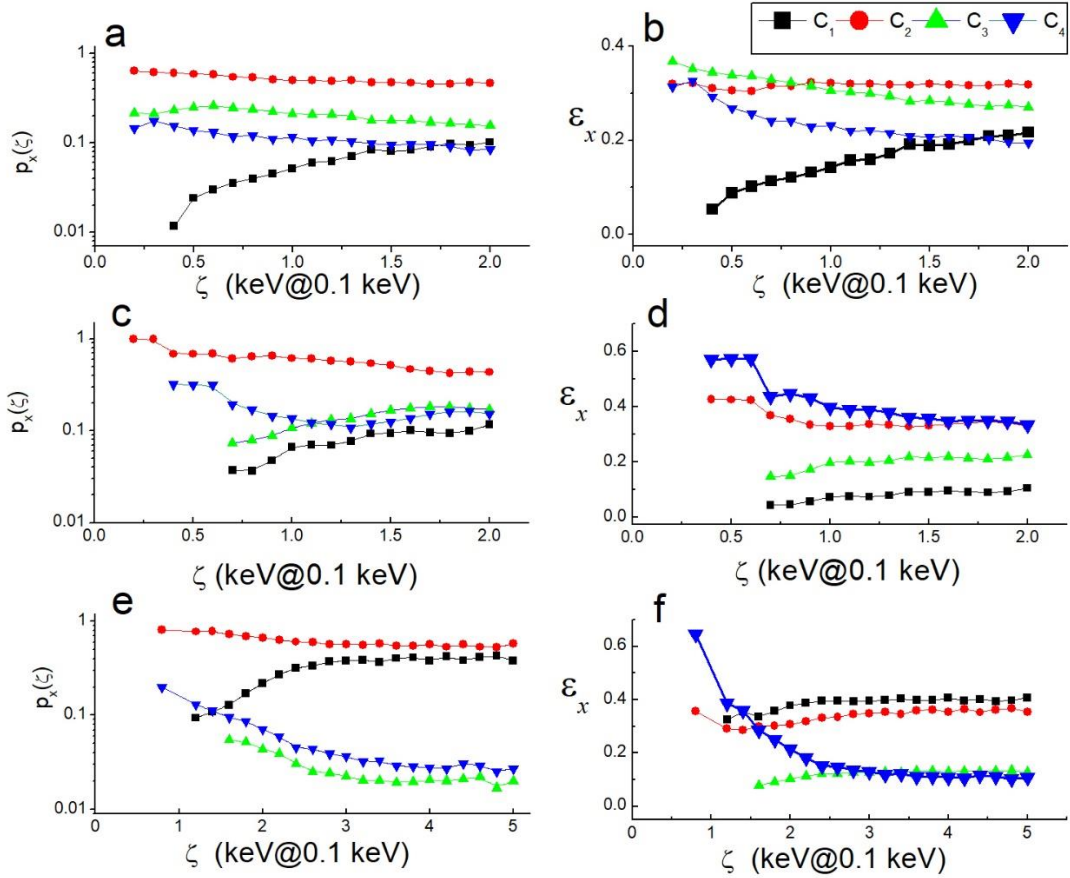

Fig. 3. Probability distributions $p_x(\zeta)$ and dynamic emergence $\varepsilon_x$ as a function of $\zeta$. (a) **SWCNTs:** $p_x(\zeta)$ as a function of $\zeta$ for $C_1^-$, $C_2^-$, $C_3^-$ and $C_4^-$ are plotted with $Cs^+$ in the energy range 0.2-2.0 keV with $\delta E(Cs^+) = 0.1\ keV$. (b) $\varepsilon_x$ versus $\zeta$ for SWCNTs show the increasing dynamic emergence of monatomic $C_1$ with $d\varepsilon_1/d\zeta > 1$, $C_2$ to $C_4$ have steady $\varepsilon_x$ profiles. (c) **C$_{60}$-fullerite:** $p_x(\zeta)$ for $C_1^-$, $C_2^-$, $C_3^-$ and $C_4^-$ for the same energy range as (a), only $C_2^-$ emitted below 0.4 keV, $C_2^-$ and $C_4^-$ up to 0.6 keV. $C_1^-$ and $C_3^-$ emerge at 0.7 keV. (d) $\varepsilon_x$ versus $\zeta$ for C$_{60}$-fullerite has distinctive emergent behavior of $C_4^-$. (e) **C$_{60}$-ZnO powder:** $p_x(\zeta)$ for $C_1^-$, $C_2^-$, $C_3^-$ and $C_4^-$ for an extended energy range 0.8 to 5.0 keV with $\delta E(Cs^+) = 0.2\ keV$. (f) Dynamic emergence $\varepsilon_x$ clearly demonstrates the diminishing profile $\varepsilon_4$ of $C_4^-$ versus $\zeta$.

$C_2$ can be seen as the most abundant sputtered component, consistently at all $E(Cs^+)$ from irradiated SWCNTs, $C_{60}$ in condensed form of fullerite and as $C_{60}$ immersed in ZnO powder. The fractal dimension $d_f^2 \sim 2$ for $C_2$ in the three experimental configurations in Fig.3.

**(i) SWCNTs:** Sputtering of $C_x; x > 1$ from irradiated SWCNTs in Fig. 3(a) demonstrates $E(Cs^+)$-independent emissions of clusters C$_2$, C$_3$ and C$_4$. However, $p_1(\zeta)$ displays $E(Cs^+)$-dependence. In Fig. 3(b), the dynamic emergence function $\varepsilon_1$ of C$_1$ displays $d\varepsilon_1/d\zeta > 0$, while those for clusters $d\varepsilon_{2,3,4}/d\zeta \sim 0$. This confirms C$_1$ as the emergent species that characterizes irradiated-SWCNTs. Emission of C$_1$ is associated with CC that lead to LTS.[7] Binary collision-based cascades are linear dissipative structures with calculated fractal dimension of C$_1$ in Fig. 3(a) is $d_f^1 \sim 1$. Fractal dimensions of C$_2$, C$_3$ and C$_4$ are $d_f^2 \sim d_f^3 \sim d_f^4 \sim 2$, evaluated from their respective $p_x(\zeta)$ in Fig. 3(a). These describe LTS as space-filling multifractal[36].

**(ii) C$_{60}$-fullerite:** C$_{60}$ as condensed fullerite irradiated with low energy $E(Cs^+)$ lead to $cage \rightarrow cage + C_x; x = 2,4$ by shrinking of the pristine C$_{60}$ cages with C$_2$ and C$_4$ emissions that generates successively increasing number densities of C$_{58}$ and C$_{56}$ in Fig. 3(c). C$_2$-only emissions at $E(Cs^+) = 0.2\ and\ 0.3\ keV$ are followed by C$_2$ and C$_4$ emissions at $E(Cs^+) = 0.4 - 0.7\ keV$. Continued irradiations lead to the shrinking cages as a $C_2 \equiv 1\ hexagon$. Icosahedral C$_{60}$ with twenty hexagons has twelve non-abutting pentagons.[42-45] The reducing number of hexagons and increasing numbers of abutting pentagons induce non-isotropic distribution of the cages' strain.[46,47] Strained cages are likely to explode when the shrinking $C_{60} - nC_2 \rightarrow C_{32}$. $C_{32}$ is claimed to be the last of the stable fullerenes.[48] For cages<$C_{32}$, explosion is most likely into fragments $C_{30} \rightarrow \sum_1^6 C_x$. The relative emission probabilities of these fragments, at the explosion stage, are according to their binding energies in the pre-explosion cages. Calculated $d_f^x$ of clusters $d_f^2 \sim d_f^3 \sim d_f^4 \sim 2$, similar to SWCNTs in Fig. 3(a). C$_1$'s fractal dimension remains $d_f^1 \sim 1$ which indicates the linear character of the mechanism responsible for its emission.

**(iii) $C_{60}$-ZnO powder:** Irradiation of $C_{60}$ in ZnO powder offers a different target constitution. It has $C_{60}$ molecules dispersed in ZnO unlike the condensed f.c.c. matrix of $C_{60}$-fullerite. Fragmentation spectra of $p_x(\zeta)$ in Fig. 3(e) is different from the one in Fig. 3(c). $C_2$ is still the dominant species while the share of $C_1$ continuously increases and $p_1(\zeta) > p_4(\zeta)$ for $E(Cs^+) = 0.8 - 2.5\ keV$. In this energy regime, $p_4(\zeta)$ consistently decreases. At $E(Cs^+) = 1.4\ keV$, $p_1(\zeta) = p_4(\zeta)$ and thereafter $p_1(\zeta) \gg p_4(\zeta)$. $C_3$ remains a minor constituent throughout the entire energy range. Here the calculated fractal dimensions of $C_1$ and $C_2$ are $d_f^1 \sim d_f^2 \sim 2$ indicate that both $C_1$ and $C_2$ are emitted from the surface of the fragmenting, shrinking cages. Dynamic emergence in Fig. 3(f) confirms the consistency of $d\varepsilon_{1,2}/d\zeta \sim 0$, whereas the rate of change of $\varepsilon_4$ for $C_4$ is negative, $d\varepsilon_4/d\zeta < 0$ implying the disappearance of cage-shrinkage route $C_x \to C_{x-4} + C_4$. The cages may be exploding when $C_4$ has become minor component with $d_f^4 \sim 1$ and $d\varepsilon_4/d\zeta \sim 0$ as is evident from figure 3(f). Dynamic emergence profile of $C_4$ indicates the onset of fragmentation→ explosion.

Supplementary material #1 includes the higher clusters ($C_5$, $C_6$) in $p_x(\zeta)$ plots as a function of $\zeta$. The emission of $C_1$, $C_3$ and $C_6$ starts at $E(Cs^+) = 0.7\ keV$, this is proposed to be initial threshold of irradiation energy for the exploding cages.

## Comparative descriptions of thermal and information-theoretic descriptions for SWCNTs and $C_{60}$ cages' fragmentation

For irradiated-SWCNTs in figure 3(a), $d_f^2 \sim 2$ implies a space-filling fractal which constitutes a localized region that emits diatomic and larger clusters when the temperature in such a region on the surface of irradiated-SWCNT $\sim T_{sub}$. We have shown, elsewhere that energies of formation of multi-vacancies is less than the energy required for a monovalent vacancy $E_{xv} < E_{1v}$.[7] For a space-filling, LTS $d_f^2 \sim d_f^3 \sim 2$. Theoretical probability of emission of a cluster $C_x$ with energy of formation $E_{xv}$ at temperature $T_{sub}$ is $p(C_x) \approx (exp(E_{xv}/T_{sub}) + 1)^{-1}$.[14] Here, $p(C_x)$ is not dependent on $E(Cs^+)$ but on $T_{sub}$ as $p_x(\zeta)$ spectra as a function of $\zeta$ in figure 3(a) confirms this conclusion. Sublimation temperature calculated from the ratios of two of the clusters are tabulated in Table 1. In figure 4(a) $p(C_2)/p(C_3)$-ratio of the probabilities of sputtered $C_2$ and $C_3$ from a LTS region on the surface of an irradiated-SWCNT is plotted as function of temperature. The consistent ratio of $p(C_2)/p(C_3) \sim 3$ in figure 3(a) yields $T_{Sub} \sim 4000K$.

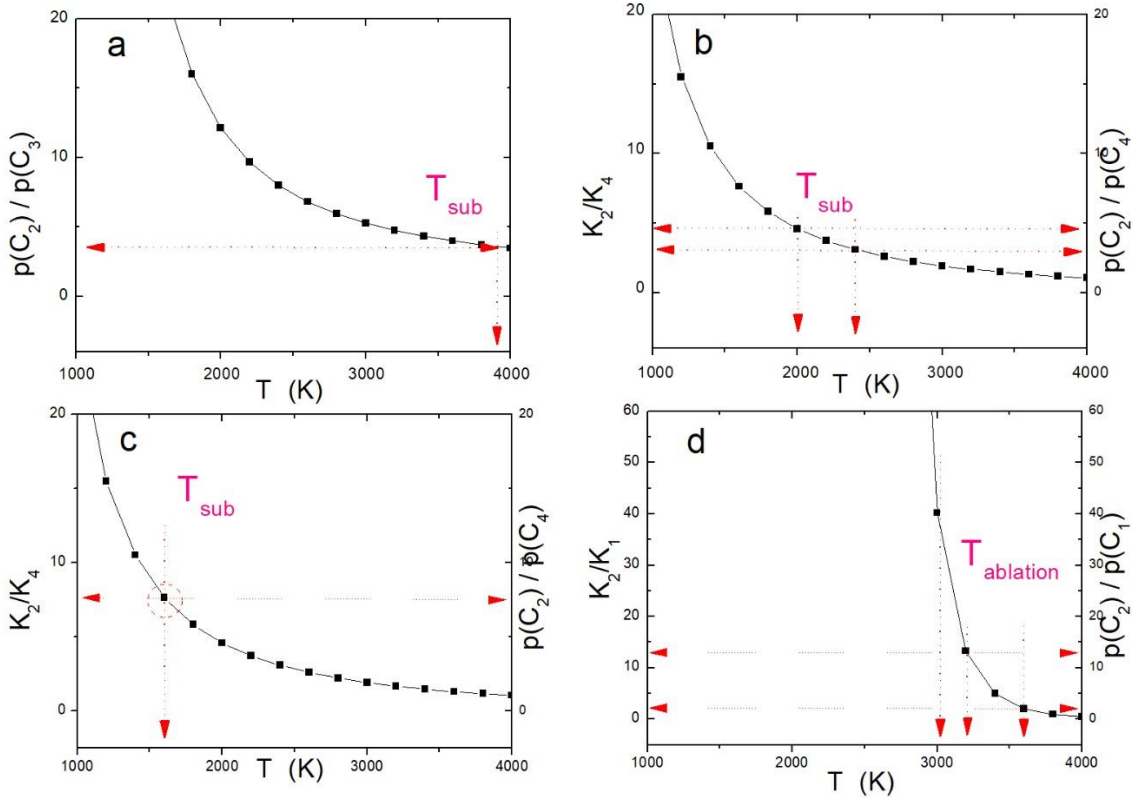

Fig. 4. Thermal and kinematical models for irradiated SWCNTs and $C_{60}$. (a) SWCNTs: Sublimation temperature $T_{Sub} \cong [(E_{xv} - E_{yv})/k)][\ln(p_y/p_x)]^{-1}$ is plotted as a function of the ratio of $C_2$ and $C_3$ from data in FIG. 3(a). The consistent ratio of $p(C_2)/p(C_3) \sim 3$ yields $T_{Sub} \sim 4000K$. (b) $C_{60}$-fullerite: $T_{Sub}$ is calculated from the rate equation $K_x =$

$Z(C_{60-x})Z(C_x)/Z(C_{60}) e^{-\frac{E(C_x)}{kT}}$ for the emissions of $C_2$ and $C_4$ from data in FIG. 3(b). Ratios of $K_2/K_4 \approx p(C_2)/p(C_4)$ are used to calculate $T_{Sub} \sim 2200 \pm 200K$. (c) $C_{60}$-ZnO: Ratios of $p(C_2)/p(C_4)$ are used to calculate $T_{Sub} \sim 1600 \pm 100K$ from plots of $K_2/K_4$ as a function of temperature. (d) $C_{60}$-ZnO: the ratio of rates $K_2/K_1$ is used to calculate $T_{Sub} \sim 3300 \pm 100K$ from experimental $p(C_2)/p(C_1)$ from FIG. 3(c).

Spherical fullerene cages of $C_{60}$ and the shrinking cages of $C_{58}, C_{56}...$, are in higher vibrational and rotational excited states as a cage. Rates of fragmentation $K_x$ of $C_{60}$ via reactions $C_{60} \rightarrow C_{60-x} + C_x$ are calculated by evaluating the partition functions of the fragmenting cage ($C_{60}$), the fragments $C_{60-x}$ and the atoms and clusters $C_x$[49]

$$K_x = \frac{Z(C_{60-x})Z(C_x)}{Z(C_{60})} e^{-\frac{E(C_x)}{kT}} \qquad (5).$$

Where partition functions of translation, rotation and vibration for the cages and the fragments are $Z(C_{60}) = z_{tr}(C_{60})z_{rot}(C_{60})z_{vib}(C_{60})$, $Z(C_{60-x}) = z_{tr}(C_{60-x})z_{rot}(C_{60-x})z_{vib}(C_{60-x})$ and $Z(C_x) = z_{tr}(C_x)z_{rot}(C_x)z_{vib}(C_x)$. The constituents of the emitted species $C_x$ are $C_1, C_2, C_3$ and $C_4$. In the first set of calculations, we have calculated $K_1$, $K_2$ and $K_4$. The ratios $K_2/K_1$ and $K_2/K_4$ are used to compare with the experimental ratios of the emitted species' probability densities $p_2(\zeta)/p_1(\zeta)$ and $p_2(\zeta)/p_4(\zeta)$. From the two sets of calculated and experimentally observed ratios, we obtain the sublimation temperature $T_{Sub}$. Supplementary Information #2 contains the details of calculations,

Table 1 compares the analytical and diagnostic capabilities of the two type of models. Thermal and the kinematical rate equation-based models yield the temperatures at which the perceived fragmentation processes operate. The inputs of thermal models are the physical parameters of cages and clusters in terms of energies of formation of vacancies in irradiated

SWCNTs and partition functions and binding energies for the rate equations applicable to $C_{60}$ cages. Thermal models can be applied where $p_x(\zeta)$ present average, steady behavior as a function of $\zeta$. This is the case for $p_2(\zeta)$, $p_3(\zeta)$ and $p_4(\zeta)$ for SWCNTs. But not for $p_1(\zeta)$ where an emergent profile of $C_1$ as a function of $\zeta$ is visible. Therefore, $C_1$ is not an output of thermal spikes in SWCNTs. In the case of fragmentation sequences demonstrated by irradiated-$C_{60}$ in figures 3(c) and (e), the emerging profile of $C_4$ is different from those of the other emissions, showing stages of the ongoing processes in fragmenting cages. Two values of $T_{sub}$ are obtained from $p(C_2)/p(C_4)$ are $T_{sub} \sim 2200K$ for $C_{60}$-fullerite and $T_{sub} \sim 1600K$ for $C_{60}$-ZnO powder. A higher value is obtained for $p(C_2)/p(C_1)$ yielding $T_{explosion} \sim 3300K$. This higher value of the temperature, does not clearly identify the underlying physical mechanisms. The plots of dynamic emergence $\varepsilon_x$ for the emitted species clarify the emergent processes as a function of $\zeta$.

A closer look at the relative entropy of $C_2$ and $C_1$ for $C_{60}$ cage fragmentation in Table 1 displays $D(p_2 \parallel p_1) = 11.0 \to 5.2 \to 0.8$ for (i) $C_{60}$-fullerite, $E(Cs^+) = 0.2 - 2.0\ keV \to$ (ii) $C_{60}$-ZnO, $E(Cs^+) = 0.8 - 2.0\ keV \to$ (iii) $C_{60}$-ZnO, $E(Cs^+) = 2.6 - 5.0\ keV$. The consistent reduction in $D(p_2 \parallel p_1)$ indicates that the emission of $C_2$ and $C_1$ describe an exploding cage. This happens when the $D(p_2 \parallel p_1) \approx D(p_1 \parallel p_2)$. The ratios of $D(p_2 \parallel p_1):D(p_1 \parallel p_2)$ are 11:7.2 for case (i) above, 5.8:5.1 for (ii) and 0.8:0.79 in the case of (iii). The cages are likely to explode when the irradiated $C_{60}$ cages are in ZnO powder, in the energy range 2.6-5.0 keV. Supplementary material #3 has Table 2 with detailed $D(p_x \parallel p_y)$ calculations.

Figure 5 presents the energy-dissipating, thermal modular descriptions of the linear-to-nonlinear transformation of $CC \to LTS$ in SWCNTs with associated time scales and fractal dimensions and the fragmenting, exploding $C_{60}$ cages. These are compare with entropy-generating, representations of the evolution and involution of dissipative structures illustrated by the information-theoretic model.

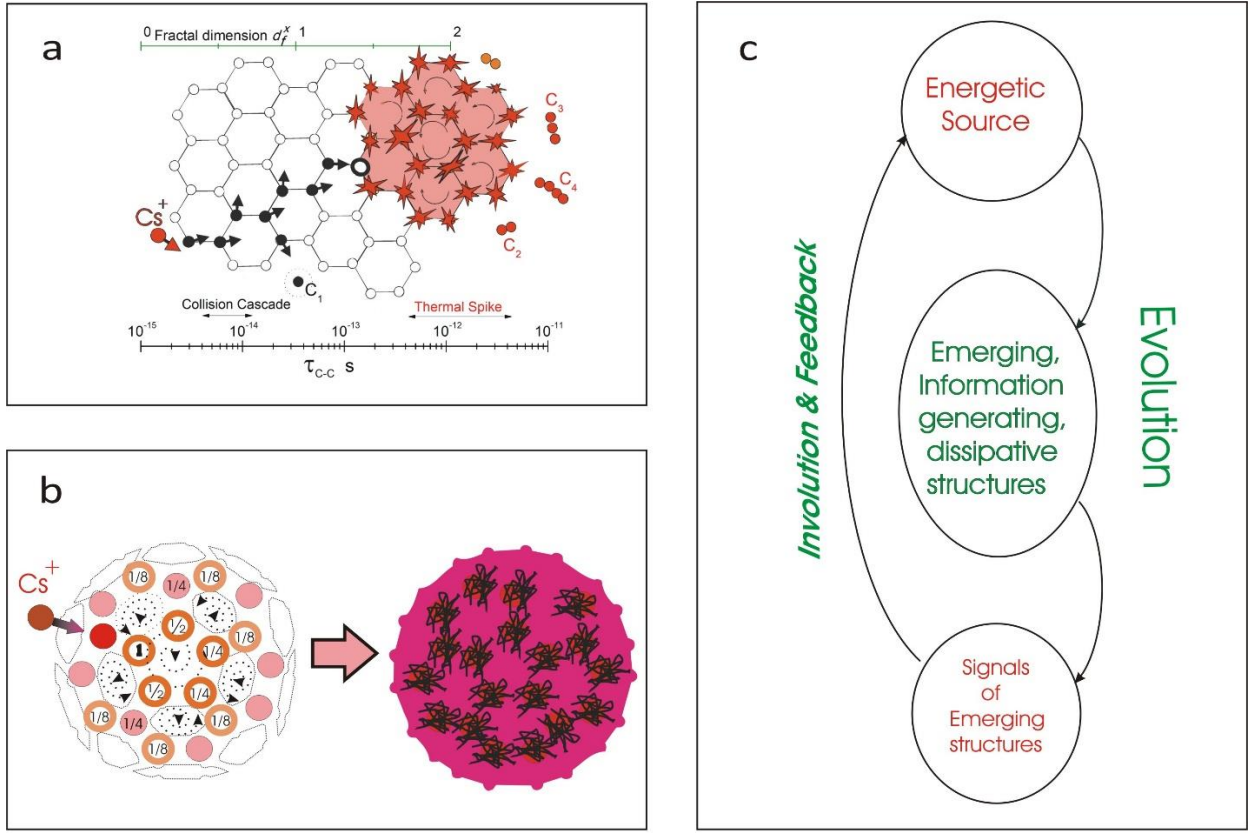

Fig. 5. Schematic representations of thermal and Inf0rmation-theoretic models. (a) In SWCNTs the initiation of collision cascades by energetic recoils is shown leading to thermal spikes that are characterized by the emission of monatomic and cluster emissions. (b) Irradiated $C_{60}$ is shown to share the primary recoil energy with all atoms that leads to hot-explosive cage. In its initial stages, only $C_2$ and $C_4$ are emitted. The shrinking cage eventually explodes. (c) The evolutionary profile of dissipative structures (cascades, spikes, fragmenting and exploding cages) in irradiated SWCNTs and $C_{60}$ cages is shown as the information-generating dynamic emergence that results from energetic input. The output signals in the form of sputtered carbon species, can be used as feedback for tailoring the emergence of the dynamical system ($Cs^+$ + C nanostructures) for desired outputs, by adjusting the input parameters by involution and feedback.

## Conclusions

In conclusion, we have presented a Source-Reservoir-Sink model for the emerging dissipative structures in irradiated SWCNTs and $C_{60}$ by utilizing the mathematical structure of Shannon's Transmitter-Channel-Receiver model for truthful transmission of input signals. However, our Source-Reservoir-Sink allow modification and manipulation of the original signal. Fractal dimension $d_f^x$ is employed as a diagnostic tool to identify the nature of the all sorts of dissipative structures that emerge due to the input energy and signal-manipulative character of the Reservoir as linear collision cascades, nonlinear localized thermal spikes in SWCNTs and fragmenting and exploding $C_{60}$ cages. Relative entropy $D(p_x \parallel p_y)$ demonstrates the information-based disparity of the nature of emerging dissipative structures. Dynamic emergence function $\varepsilon_x$ tracks the evolving and diminishing profiles of ion-induced dissipative structures. Thermal and kinematical models yield relevant temperatures of fragmenting-to-exploding cages based on assumptions like binding energies, partition functions etc. The information-theoretic functions $I_x$, $d_f^x$, $D(p_x \parallel p_y)$ and $\varepsilon_x$ are independent of any preconceived physical processes and *apriori* assumptions. These are based on the calculations of the experimental probability distribution functions of the emitted species that are treated as signals of the evolving, self-organizing dissipative structures.

## Acknowledgements

Author is indebted to S. Zeeshan, S. Javeed, S. A. Janjua, S. D. Khan. A. Ashaf, K. Yaqub, M. Khalil and M. Yousuf for participation and support in numerous experiments reported over the last decade that form the bulk of data referred to. The SRS model employed here was presented in the extended discourse at PIEAS, Islamabad, where the author (SA) gave a course 'Fractals and Irradiated Solids' in spring of 2018. S. M. Mirza, M. Y. Hamza, S. Qamar, M. Ikram and M. T. Siddiqui and colleagues and students at PIEAS are acknowledged for their support and helpful discussions during the course.

## Additional Information

I declare that the authors have no competing interests as defined by Nature Research, or other interests that might be perceived to influence the results and/or discussion reported in this paper.

Table 1. Comparison of the physical parameters derived from thermal and kinematical models with the diagnostic tools derived from Information-theoretic model

| $Cs^+$-irradiated Carbon Nanostructure | Thermal and Kinematical Models $T_{Sub}$ and $T_{explosion}$ | Information-theoretic Model $d_f^x$, $D(p_x \parallel p_y)$ and $d\varepsilon_x/d\zeta$ |
|---|---|---|
| **SWCNT** $E(Cs^+) = 0.2$ to 2.0 keV | $p_x = \{exp(E_{xv}/kT_S) + 1\}^{-1}$ $T_{Sub} \cong [(E_{xv} - E_{yv})/k)][\ln(p_y/p_x)]^{-1}$ $p(C_2)/p(C_3) \to T_{sub} \cong 4000 \pm 200\ K$ | $d_f^2 \sim d_f^4 \sim 2$ $d_f^1 \sim 1$ $D(C_2 \parallel C_1) = 14.5$ $D(C_2 \parallel C_4) = 9.6$ $\boldsymbol{d\varepsilon_1/d\zeta > 1}$ |
| **$C_{60}$-Fullerite** $E(Cs^+) = 0.2$ to 2.0 keV | $K_x = Z(C_{60-x})Z(C_x)/Z(C_{60})\ e^{-\frac{E(C_x)}{kT}}$ $\frac{K_2}{K_4} \approx p(C_2)/p(C_4) \to T_{sub}$ $\cong 2200 \pm 200\ K$ | $d_f^2 \sim d_f^4 \sim 2$ $d_f^1 \sim 1$ $D(C_2 \parallel C_1) = 11.0$ $D(C_2 \parallel C_4) = 7.1$ $\boldsymbol{d\varepsilon_{2,4}/d\zeta \sim 0}$ |
| **$C_{60}$ in ZnO powder** $E(Cs^+)$ (i) 0.8 to 2.6 keV & | $K_x = Z(C_{60-x})Z(C_x)/Z(C_{60})\ e^{-\frac{E(C_x)}{kT}}$ $K_2/K_4 \approx p(C_2)/p(C_4) \to T_{sub}$ $\cong 1600 \pm 100\ K$ | $d_f^1 \sim d_f^2 \sim 2$ $d_f^4 \sim 1$ $D(C_2 \parallel C_1) = 5.2\ \&\ 0.8$ $D(C_2 \parallel C_4) = 10.1\ \&\ 16.4$ |

| (ii) 2.6 to 5.0 keV | $K_2/K_1 \approx p(C_2)/p(C_1) \to T_{explosion}$ $\cong 3300 \pm 100\ K$ | $d\varepsilon_4/d\zeta < 1$ |